\documentclass{article}
\usepackage{spconf,amsmath,graphicx}

\title{Inclusive Speaker Verification with Adaptive Thresholding}
\name{Navdeep Jain, Hongcheng Wang}
\address{
  Comcast Applied AI Research 
  }

\begin{document}
%
\maketitle
\begin{abstract}
While using a speaker verification (SV) based system in a commercial application, it is important that customers have an inclusive experience irrespective of their gender, age, or ethnicity. In this paper, we analyze the impact of gender and age on SV and find that for a desired common False Acceptance Rate (FAR) across different gender and age groups, the False Rejection Rate (FRR) is different for different gender and age groups. To optimize FRR for all users for a desired FAR, we propose a context (e.g. gender, age) adaptive thresholding framework for SV. The context can be available as prior information for many practical applications. We also propose a concatenated gender/age detection model to algorithmically derive the context in absence of such prior information. We experimentally show that our context-adaptive thresholding method is effective in building a more efficient inclusive SV system. Specifically, we show that we can reduce FRR for specific gender for a desired FAR on voxceleb1 test set by using gender-specific thresholds. Similar analysis on OGI kids' speech corpus shows that by using age specific threshold, we can significantly reduce FRR for certain age groups for desired FAR.    
\end{abstract}
\section{Introduction}
\label{sec:intro}

SV can be used to verify individuals for secure, frictionless user engagements in a wide range of solutions such as customer identity verification in call centers, contact-less facility access, personalized service offerings on smart speakers or tv. According to different application scenarios, SV can be categorized as text-dependent (TD-SV) and text-independent (TI-SV). For TD-SV, the spoken content of the test utterance and the enrollment utterance should be the same, whereas there is no constraint on the spoken content in the TI-SV system.

In recent years, publicly available free large datasets such as voxceleb1~\cite{Nagrani-2017} and voxceleb2~\cite{Chung-2018} have accelerated deep learning-based model training for SV tasks. ResNet~\cite{He-2015} based backbone networks have shown promising results for this work~\cite{ Xie-2019, Zeinali-2019, Zhou-2019}. The accuracy of ResNet-based networks can further be improved by increasing the depth and width of a network. However, when input feature space becomes more complicated just increasing dimensions of the ResNet network leads to overfitting.  Several changes have been proposed to further improve the performance of the ResNet such as ResNeXt~\cite{Xie-2017} and Squeeze and Excitation (SE) network,~\cite{Hu-2018}. The ResNeXt block replaces the residual block in the ResNet with a multi-branch transformation by introducing many groups of convolution in one layer. The SE block adaptively recalibrates channel-wise feature responses by explicitly modeling interdependencies between channels. Another such enhancement is Res2Net~\cite{Gao-2019}. Res2Net adds two additional dimensions to Resnet, cardinality, and scale. Res2Net constructs hierarchical residual-like connections inside the residual block and assembles variable-size receptive fields within one layer. Res2Net has shown promising results for SV tasks in recent work~\cite{Zhou-2020}.

Similar to network architectures, multiple aggregation mechanisms have been proposed to generate embedding to further improve the performance of algorithm~\cite{Zhu-2018, Okabe-2018, Kye-2020}. Different loss functions have been proposed to optimize network performance for SV tasks~\cite{Mingote-2019, Chung-2020}. Among all these advancements, one of the under-explored aspects is how natural attributes of speakers, such as gender, age, accent affects model accuracy and how the model accuracy can be improved if the attributes are provided either as prior information or can be inferred. 

\begin{figure*}[t]
  \centering
  \includegraphics[width=0.7\textwidth]{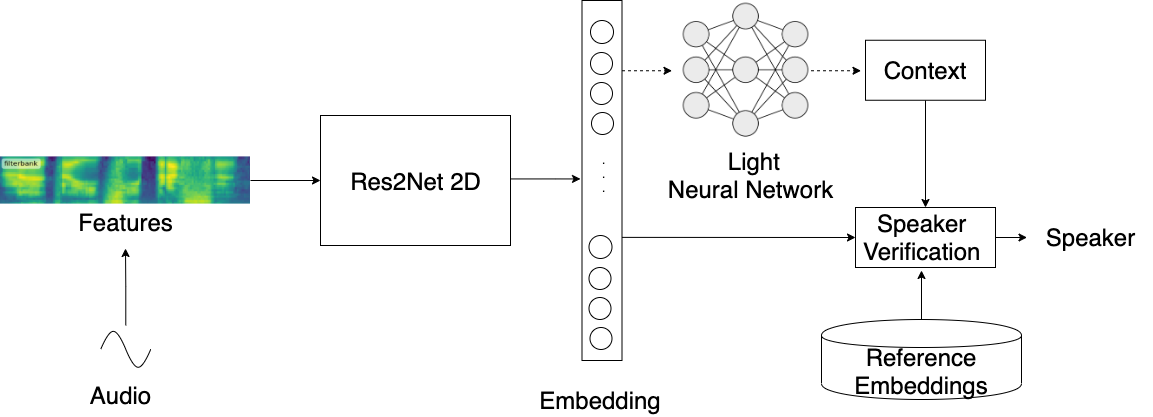}
  \caption{Speaker Verification with context-adaptive thresholding. A trained Res2Net 2D-based neural network generates speaker embedding.  Optionally, a trained fully connected light neural network detects context (such as gender, age or ethnicity). The Speaker verification process selects the correct threshold based on context to verify the speaker.}
  \label{fig:Concatenated Detector}
\end{figure*}

In this paper, we analyze how gender and age affect SV using our state-of-the-art speaker embedding model. We perform this analysis using two test datasets. The first dataset is voxceleb1 test set. We use this test set to analyze model performance on gender. The second dataset is OGI kids' speech corpus~\cite{Shobaki-2000}. We use the second test set to analyze how the model performs for different age groups of kids. Our analysis shows that by using different thresholds for different gender and different age groups, we can achieve desired FAR goal for all groups with reduced FRR for some of the groups as compared to using a single threshold for all the data. To make use of such adaptive thresholds, details of the speaker's gender and age are known or can be inferred. Many verification/identification applications such as banking or customer care call center have the customer's personal information, so such information can be leveraged as prior knowledge. In absence of availability of such information, such as TV entertainment or smart speaker use case, where you may have multiple family members using the same voice remote control or speaker, we propose to infer the gender/age information as a context for SV, by training a separate gender/age classifier using the output of speaker embedding model as input to such classifier.

The rest of the paper is organized as follows: Section 2 describes our context based SV approach. Experimental results are presented and discussed in Section 3. We describe the conclusion and future work in Section 4.

\section{Context based Speaker Verification}
\label{sec:Context based Speaker Verification}
Fig.~\ref{fig:Concatenated  Detector} summarizes our approach for SV with context-adaptive thresholding. The 80-dimensional kaldi filterbank~\cite{Ravanelli-2018} is extracted as features from the input audio. A Res2Net 2D building block based neural network model is trained to generate highly distinguishable embedding for different speakers using features. This model has few million trainable parameters. If details such as the speaker's gender, age, and accent are available, the SV process will use those details to pick up the correct threshold to make a verification decision. If such information is not available, we propose training light neural networks to classify such details. A light neural network is a  small fully connected neural network with few thousand trainable parameters. Light neural networks are trained using embedding generated using the Res2Net-based trained model. Due to the smaller size of the light neural networks, they can be used along with the speaker embedding model during the verification process without any significant additional load on GPU or latency in the process.

\subsection{Res2Net based Speaker Embedding Model}
\label{ssec:Res2Net  based  Speaker Embedding Model}
This section describes the structure of the Res2Net-based text-independent speaker  embedding model developed for this study. A typical end-to-end speaker embedding system consists of three main components. First of all, a frame-level speaker embedding extractor. Then, an aggregation layer is applied to summarize the frame-level representations and yield a fixed-dimensional utterance-level speaker embedding. Lastly, a speaker-discriminative loss function is used to minimize the training objective. We describe the details of these components in the remaining of this section. Complete model architecture is summarized in Table~\ref{table:Res2Net Architecture}. 

\subsubsection{Res2Net  Module}
\label{sssec:Res2Net Module}
\begin{figure}[h]
  \centering
  \includegraphics[width=\linewidth]{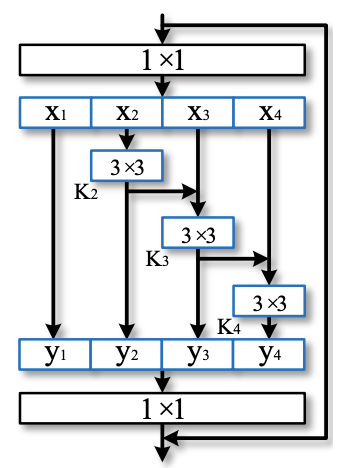}
  \caption{Res2Net Module (scale=4)}
  \label{fig:Res2Net Module}
\end{figure}
Res2Net is used as a building block of the network. Fig.~\ref{fig:Res2Net Module} illustrates the design of Res2Net module. After going through the first 1×1 layer, feature maps are evenly sliced into s subsets, denoted by {x1, x2, ..., xs}. Each subset is then fed into a 3×3 convolution (denoted by $K_i$), except for x1. Starting from x3, each output of $K_i$-1 is added with xi before going through $K_i$. This hierarchical residual-like connection further increases the possible receptive fields within one layer, resulting in multiple feature scales. The whole process can be formulated as
\begin{equation}
y_i =
\begin{cases}
 x_i, &  i = 1;\\
 K_i(x_i), & i=2;\\
 K_i(x_i+y_{(i-1)}), & i=3,4,...,s\\
\end{cases}
  \label{eq1}
\end{equation}
where {$y_1, y_2, ..., y_s$} is the output of this module, they are concatenated and fed into the following 1 × 1 convolutional layer to maintain the channel size of this residual block. With the scale dimension, the Res2Net model can represent multi-scale features with various granularity, which facilitates SV for very short utterances. 

\begin{table}[h!]
\caption{Speaker Embedding Model Network Architecture. For Res2Net, basewidth w is 26 and scale is 8. The embedding size is 256.}.
\centering
\begin{tabular}{c c} 
 \hline
 Layer Name & Structure  \\ [0.5ex] 
 \hline
  Input &  80-dimensional Kaldi filterbank\\
  Conv2D & $7\times7$, stride (2,1), padding 3  \\ [0.5ex] 
 \hline
 \\
 Res2NetBlock-1 & $\begin{bmatrix}
 1\times1 & 64\\ 
 3\times3& 64\\ 
 1\times1& 256
\end{bmatrix} \times 3$, stride=1  \\ 
\\
 Res2NetBlock-2 & $\begin{bmatrix}
 1\times1 & 64\\ 
 3\times3& 64\\ 
 1\times1& 256
\end{bmatrix} \times 4$, stride=2 \\
\\
 Res2NetBlock-3 & $\begin{bmatrix}
 1\times1 & 128\\ 
 3\times3& 128\\ 
 1\times1& 512
\end{bmatrix} \times 6$, stride=2  \\
\\
 Res2NetBlock-4 & $\begin{bmatrix}
 1\times1 & 128\\ 
 3\times3& 128\\ 
 1\times1& 512
\end{bmatrix} \times 3$, stride=1  \\ [1ex] 
\\
\hline
 Average Pooling & pooling in frequncey domain\\
 Aggregatoin & ASP \\
\hline
 Dense1 & embedding size\\
 Dense2 & AM-Softmax \\
\hline
\end{tabular}
\label{table:Res2Net Architecture}
\end{table}

\subsubsection{Aggregation}
\label{sssec:Aggregation}

After the Res2Net layer, average pooling is applied in the frequency domain and then  Attentive Statistics Pooling (ASP)~\cite{Okabe-2018} is used to generate fixed-length speaker embedding. ASP utilizes an attention mechanism to give different weights to different frames and generates not only weighted means but also weighted standard deviations. In this way, it can capture long-term variations in speaker characteristics more effectively.

\subsubsection{Loss Function}
\label{sssec:Loss Function}
Additive Margin version of SoftMax (AM-Softmax)~\cite{Wang-2018} is used as a loss function. AM-Softmax introduces a concept of a margin between classes to increase inter-class variance. We select a margin of 0.2 and a scale of 30 since these values give the best results on the VoxCeleb1 test set.

\subsection{A Light Neural Network for Gender Detection}
\label{ssec:A Light Neural Network for Gender Detection}
\begin{table}[h!]
\caption{Gender Detector Network Architecture . Embedding size is 256 and no.of classes is 2}.
\centering
\begin{tabular}{c c} 
 \hline
 Layer Name & Structure  \\ [0.5ex] 
 \hline
  Input &  embedding size \\
 \hline

\hline
 Dense1 & 128\\
 Dense2 & 256\\
 Dense3 & no. of classes\\
\hline
Loss & Cross-Entropy\\  
\hline
\end{tabular}
\label{table:Light Neural Network Architecture}
\end{table}

Even though our analysis focuses on prior knowledge of contexts such as gender and age, in many applications, the context may not be available. To address this issue, we propose a novel scheme in which embedding generated by a trained speaker embedding model is given as input to light fully connected neural networks for context detection. Though the speaker embedding is learned to discriminate different speakers, the embedding has intrinsically encoded the speaker's many attributes, including gender, age, and accent. Therefore we do not need to build a complicated model to classify such attributes, instead, we leverage this embedding combined with light neural networks for context detection.  The whole pipeline becomes much more efficient compared to building a complicated context classification model using raw audio as input. Due to widely available labeled public data, we only focus on gender detection using such an approach in this study. We train a fully connected neural network for gender detection. Network architecture for gender detector is summarized in Table~\ref{table:Light Neural Network Architecture}. This gender detector has only 67,716 trainable parameters. While evaluating, as shown in Fig.~\ref{fig:Concatenated Detector}, the speaker embedding model generates embedding from audio features, and this embedding is passed to the gender detector as input features. The gender output is then used to select the optimal threshold to make the SV decision.   

\section{Experiments}
\label{sec:Experiments}

\subsection{Speaker Embedding Model Training and Performance}
\subsubsection{Training Data}
Following 3 datasets are used for model training : 1) Voxceleb2~\cite{Chung-2018} dev set – 5994 speakers, 2) Voxceleb1~\cite{Nagrani-2017} dev set – 1211 speakers, 3) CSTR VCTK corpus~\cite{Yamagishi-2019} – 110 speakers. This results into total of 7315 speakers.

\subsubsection{Data Augmentation}

Augmentation increases the amount and diversity of the training data, which helps reduce over-fitting. We employ two of the popular augmentation methods in speech processing – additive noise and room impulse response (RIR) simulation. For additive noise, we use the audio clips from the MUSAN corpus~\cite{Snyder-2015}. For RIR, we sample the simulated filters of small and medium rooms released in~\cite{Ko-2017}. Kaldi recipe is used for data augmentation.

\subsubsection{Input Features}
Features are extracted by randomly selecting 2 seconds of audio from a segment. The 80-dimensional kaldi filterbank~\cite{Ravanelli-2018}  are used as features with a frame-length of 25 ms and a frame-shift of 10 ms.

\subsubsection{Model Performance}
Model performance on voxceleb1 test set is summarized in Table~\ref{Res2Net Model Performance}. Audio Embedding is generated by processing the full length of audio. Cosine similarity scoring is used on the embedding without any prepossessing. Equal Error Rate (EER)  and minimum detection cost function (minDCF) are used as metrics to measure performance. Trained model has an EER of 0.82\% and minDCF of 0.047 and 0.076 for  p\_target of 0.05 and 0.01 respectively. The model has 7.04M parameters and 8G multiply-accumulate operations (MACs). Our results are on par with winners of voxsrc 20~\cite{Nagrani-2020} for a single, non-fusion, model on the same test set.

\begin{table}[h!]
\caption{Result of the system on the voxceleb1 test set.  minDCF values are for p\_target 0.05 and 0.01}. 
\centering
\begin{tabular}{*4c } 
 \hline
 EER(\%) & minDCF  & no. of parameters & MACs\\ 
 \hline
0.82 & 0.047 / 0.076 & 7.04 M  & 8 G \\
\hline
\end{tabular}
\label{Res2Net Model Performance}
\end{table}

\subsection{Gender detector Training and Performance}
\label{ssec:Gender detector Training and Performance}
\begin{figure}[h]
  \centering
  \includegraphics[width=\linewidth]{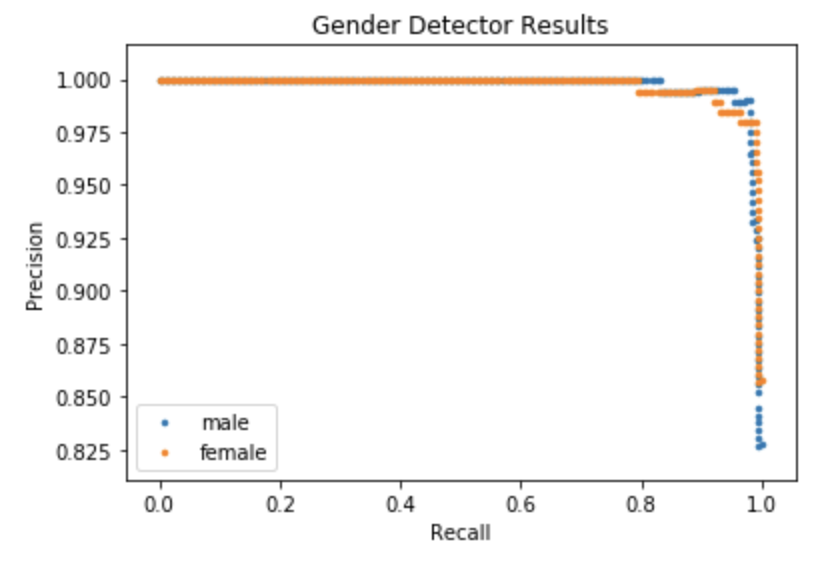}
  \caption{Precision-Recall Curve for Gender Detector}
  \label{fig:Precision-Recall Curve for Gender Detector)}
\end{figure}

To have samples from a large number of speakers of different genders, for gender detector training, we use data of male and female speech from the training set of google audio set~\cite{45857} and vgssound~\cite{Chen-2020}. Training data is comprised of 19726 male speech 10477 female speech. Testing data is comprised of evaluation and test sets of 197 male speech and 199 female speech from google audio set. Both training and test full-length audio are passed through the trained speaker embedding model to generate input features for the gender detector.  The Precision-Recall curve of the trained gender detector on the test set is shown in Fig.~\ref{fig:Precision-Recall Curve for Gender Detector)}. 

As you can see in Fig.~\ref{fig:Precision-Recall Curve for Gender Detector)}, even a small model with few thousand parameters has a very high accuracy of classification on the test set. The same model identifies the gender of all 40 speakers correctly for all 4715 unique audio samples in voxceleb1 test set.

\subsection{Threshold Selection Analysis for Gender}
\label{ssec:Threshold Selection Analysis on Gender}
\subsubsection{Test sets}
\label{sssec:Gender Test Sets}
For gender-specific analysis, we use voxceleb1 test set. It has a total of 40 speakers: 25 male, 15 female. The default test set has a total of 37720 pairs. We present our analysis using 2 subsets of this full test set. Test set 1 contains all the 29612 same-gender pairs: 22576 male speaker pairs, 7036 female speaker pairs. To make sure, imbalance in gender-wise count as well per gender pair count doesn't affect the result, we also present results for the balanced test set, test set 2, where we randomly pick 15 male speakers out of 25 and 7036 pairs of those 15 male speakers, so we have an equal number of speakers per gender i.e. 15 and an equal number of pair per gender i.e. 7036.

\subsubsection{Result and Analysis}
\label{sssec:Gender Result and Analysis}
\begin{figure}[h]
  \centering
  \includegraphics[width=\linewidth]{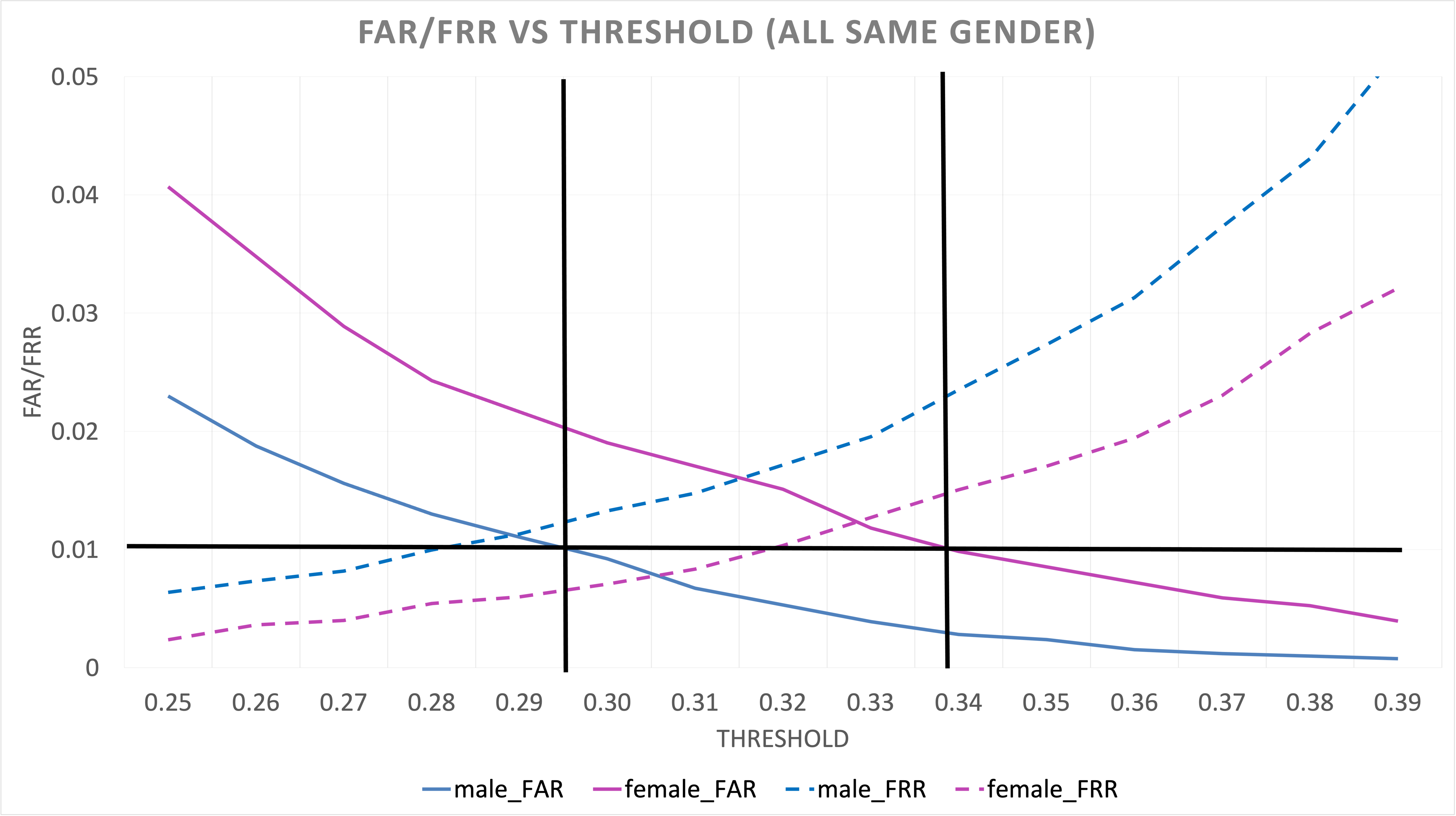}
  \caption{FAR/FRR on the test set 1 (all same-gender pairs). Male FAR is 0.01 at close to threshold 0.3, Female FAR is 0.01 at close to threshold 0.34. Male FRR increases from 0.013 at threshold 0.3 to 0.023 at a threshold of 0.34}
  \label{fig:FAR/FRR on test set 1 (all same gender pairs)}
\end{figure}

\begin{figure}[h]
  \centering
  \includegraphics[width=\linewidth]{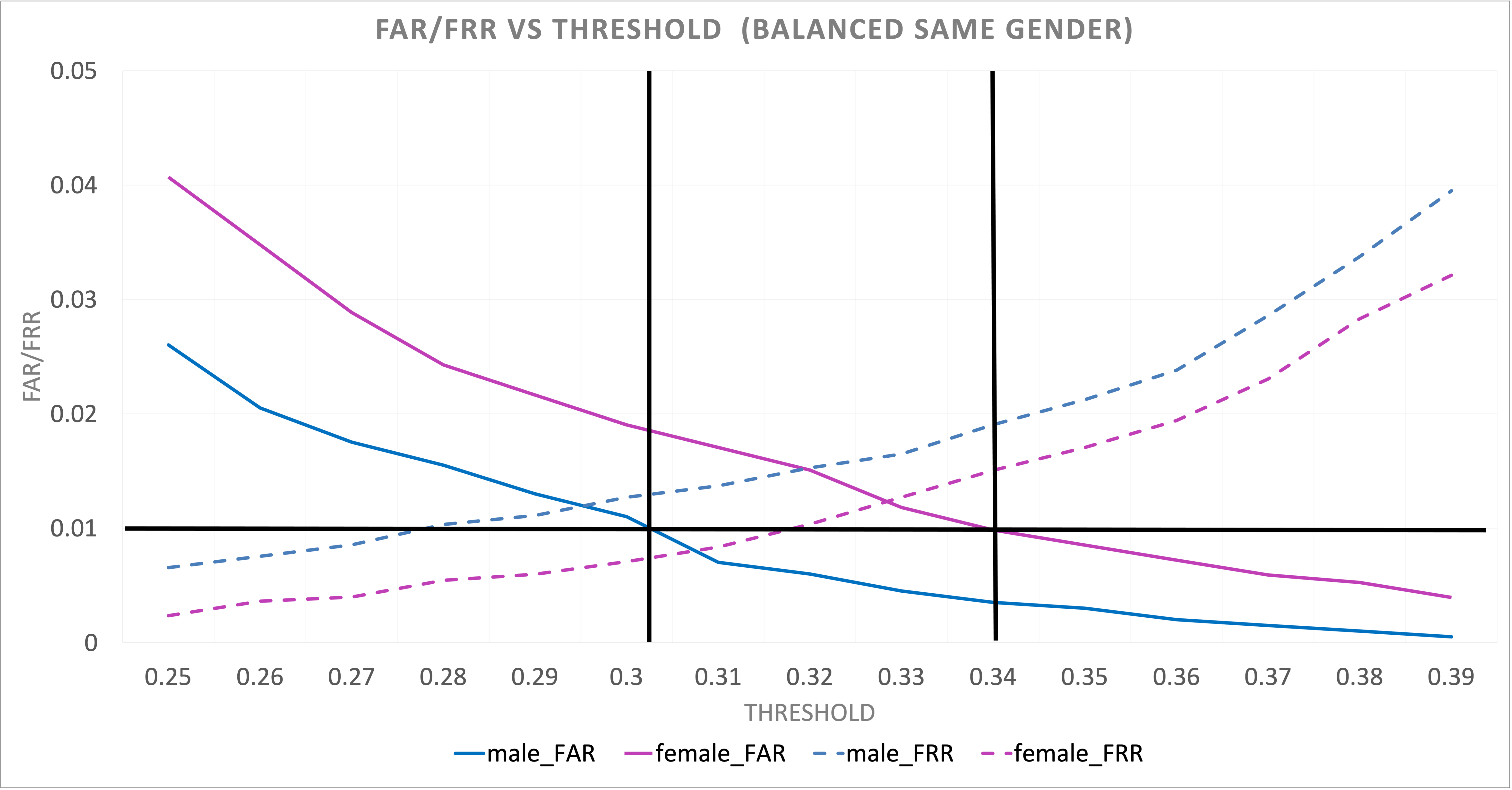}
  \caption{FAR/FRR on the test set 2 (balanced same-gender pairs). Male FAR is 0.01 at close to  threshold  0.3, Female FAR is 0.01 at close to threshold 0.34. Male FRR increases  from 0.012 at threshold 0.3 to 0.019 at a threshold of 0.34 }
  \label{fig:FAR/FRR on test set 2 (balanced same gender pairs) }
\end{figure}
\begin{table*}[h]
\caption{FAR/FRR summary for different test sets for state-of-the-art SV model}.
\centering
\begin{tabular}{*8c} 
 \hline
\multicolumn{2}{c}{test set} & 0.01 FAR  & 0.01 FAR  &\multicolumn{2}{c}{male FRR } &\multicolumn{2}{c}{male FRR }\\ 
\multicolumn{2}{c}{}&  threshold & threshold & \multicolumn{2}{c}{at male 0.01 FAR }& \multicolumn{2}{c}{at female 0.01 FAR }\\

\multicolumn{2}{c}{}&  male &  female & \multicolumn{2}{c}{threshold}& \multicolumn{2}{c}{threshold}\\

 \hline
\multicolumn{2}{c}{all same-gender pairs} &0.3 &0.34 &\multicolumn{2}{c}{0.013} &\multicolumn{2}{c}{0.023}\\
\multicolumn{2}{c}{balanced same-gender pairs} & 0.3 &0.34 &\multicolumn{2}{c}{0.012} &\multicolumn{2}{c}{0.019} \\
\hline
\end{tabular}
\label{table:imbalanced training FAR/FRR}
\end{table*}
Fig.~\ref{fig:FAR/FRR on test set 1 (all same gender pairs)} and Fig.~\ref{fig:FAR/FRR on test set 2 (balanced same gender pairs) } show FAR/FRR vs threshold curves for test set 1 and 2 respectively. To highlight the point of picking up gender-specific thresholds, let's assume that the FAR target for application is 0.01. For this FAR, result is  summarized in Table~\ref{table:imbalanced training FAR/FRR}. As you can see for both test sets, for females, desired FAR is achieved at higher thresholds as compared to thresholds at which desired FAR is achieved for males. If we use one threshold, the threshold at which female FAR is 0.01 then male FRR increases by 77\%  and 58\% for test set 1 and 2 respectively, as compared to the threshold at which male FAR is 0.01. Thus, by using gender-specific thresholds we can reduce FRR for male and broadly overall data while achieving desired FAR goal.

\subsubsection{Impact of imbalanced number of speakers per gender in training}
\label{sssec:Impact of imbalanced number of speakers per gender in training}
Our training set for the state-of-the-art model has almost 59\% of male speakers and 31\% of female speakers. To eliminate the impact of the imbalanced number of speakers per gender, we also train a model with an equal number of female and male speakers. This second model is trained with 2858 speakers of each gender. Model architecture remains exactly the same as described in section~\ref{ssec:Res2Net  based  Speaker Embedding Model}. Due to reduced training data, EER for the second model on voxceleb1 test set is 1.2. 

\begin{table*}[h]
\caption{FAR/FRR summary for different test sets for a model trained with an equal number of per gender speakers}.
\centering
\begin{tabular}{*8c} 
 \hline
\multicolumn{2}{c}{test set} & 0.01 FAR  & 0.01 FAR  &\multicolumn{2}{c}{male FRR } &\multicolumn{2}{c}{male FRR }\\ 
\multicolumn{2}{c}{}&  threshold & threshold & \multicolumn{2}{c}{at male 0.01 FAR }& \multicolumn{2}{c}{at female 0.01 FAR }\\

\multicolumn{2}{c}{}&  male &  female & \multicolumn{2}{c}{threshold}& \multicolumn{2}{c}{threshold}\\

 \hline
\multicolumn{2}{c}{all same-gender pairs} &0.31 &0.34 &\multicolumn{2}{c}{0.021} &\multicolumn{2}{c}{0.034}\\
\multicolumn{2}{c}{balanced same-gender pairs} & 0.31 &0.34 &\multicolumn{2}{c}{0.023} &\multicolumn{2}{c}{0.038} \\
\hline
\end{tabular}
\label{table:balanced training FAR/FRR}
\end{table*}
The performance of this second model on test sets described in section~\ref{sssec:Gender Test Sets} is summarized in Table~\ref{table:balanced training FAR/FRR}. As you can see, even after using an equal number of speakers per gender, the threshold at which we achieve specific FAR, in our case 0.01, varies for different gender. By using gender-specific thresholds, rather than a single threshold - the threshold at which all gender have desired FAR, we can optimize FRR for all genders. As described in the Table~\ref{table:balanced training FAR/FRR}, male FRR increases by 61.9\% and 65.2\% for test set 1 and 2 respectively while using a single threshold, the threshold at which female FAR is 0.01 vs using threshold at which male FAR is 0.01. These results align with results shared in section~\ref{sssec:Gender Result and Analysis}. 

Thus, whether we use an equal number of speakers per gender in training and/or testing, thresholds at which FRR optimizes for a desired common FAR vary for different gender.

\subsection{Threshold Selection Analysis for Age}
\label{ssec:Threshold Selection Analysis on Age}
\subsubsection{Test sets} 
A subset of OGI kids' speech corpus~\cite{Shobaki-2000} is used for age-specific analysis. 10 kids are randomly picked from each grade between 1-10. Thus, the test set has a total of 100 kids. Moreover, kids are divided into three groups. Group1 has kids from grades 1-5, group2 has kids from grades 6-7 and group3 has kids from grades 8-10. These groups are chosen based on EER thresholds (EET) for each grade. EET for every grade in a particular group is close to each other. In total, we have 27659 pairs in the test set.

\subsubsection{Result and Analysis}
\begin{figure}[h]
  \centering
  \includegraphics[width=\linewidth]{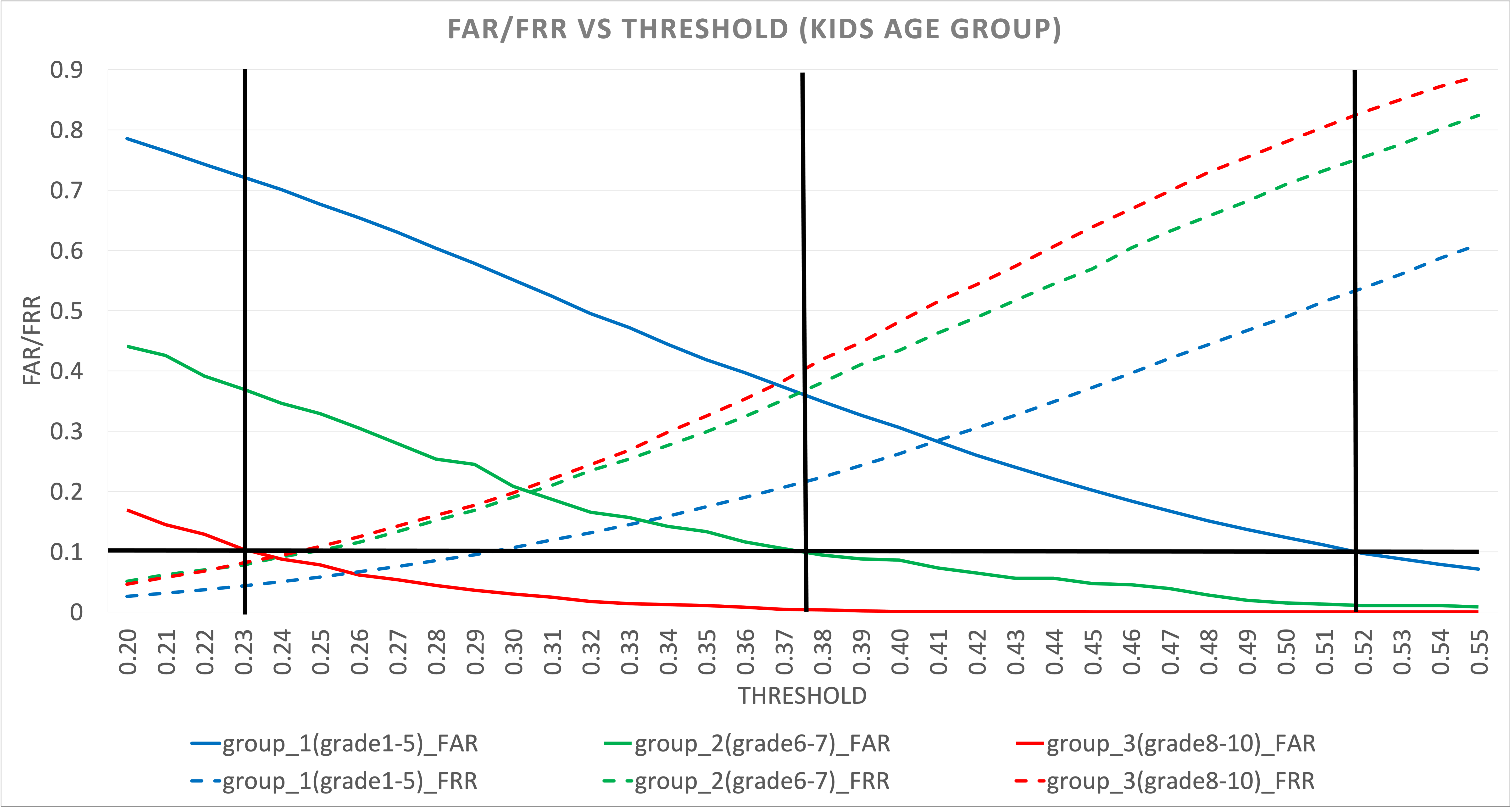}
  \caption{FAR/FRR for kids age groups. Group 1, 2, and 3 have FAR of 0.1 at thresholds 0.52, 0.38, and 0.23 respectively. For the group3, FRR increases from 0.08 at threshold 0.24 to 0.83 at a threshold of 0.52. For the group2, FRR increases from 0.38 at threshold 0.38 to 0.75 at a threshold of 0.52}
  \label{fig:FAR/FRR on Kids)}
\end{figure}
Fig.~\ref{fig:FAR/FRR on Kids)} shows results for different age groups of kids. For this application, say desired FAR is 0.1. As kids' data isn't used in training, the error tends to stay higher for kids speakers compared to error for adult speakers. For group1, group2, and group3, this FAR can be achieved at threshold 0.52, 0.38, and 0.23 respectively. If we use age-group specific thresholds, we can reduce FRR for group2 and group3 by close to 50\% and 90\% respectively compared to using a single threshold, the threshold at which group1 and in turn, all groups have desired FAR.

\section{Conclusions}
Our analysis shows that rather than using a single threshold for speakers of different gender if we have prior knowledge of the gender of speakers and we use gender-specific thresholds for different gender, for a maximum acceptable FAR for all genders, we can optimize FRR for all genders. Our analysis also shows that this result stands even if there is same number of speakers per gender in training and/or test data. Our analysis of kids of different age groups shows a similar result. Rather than using a single threshold for different age groups of kids if we use age group-specific thresholds, we can reduce FRR for certain age groups for a desired FAR. Thus, by using adaptive thresholds on basis of context such as gender, age, we can achieve inclusive FAR target for all groups while minimizing  FRR for respective groups.

This approach requires prior knowledge of the gender and age of speakers. Even though we believe that it is feasible to have such details in advance for many SV use cases such as banking call centers, there are use cases where such metadata may not be available in advance. In absence of prior knowledge of such information, we also propose training a simple fully connected neural network using embedding generated from trained speaker embedding model as input features to this simple neural network. Our results on gender classifier with the above approach show that we can train a highly accurate model to detect gender from a speech. Also, a classifier trained in such a manner has very low latency and adds very little delay in the overall SV process.

For future work, we would like to analyze, how age affects SV of adult males and females. We'll also evaluate whether we can train a similarly accurate light neural  network for age  group detection.

\bibliographystyle{IEEEbib}
\bibliography{refs}

\begin{thebibliography}{10}

\bibitem{Nagrani-2017}
A.~Nagrani, J.~S. Chung, and A.~Zisserman,
\newblock ``Voxceleb: a large-scale speaker identification dataset,''
\newblock in {\em Interspeech}, 2017.

\bibitem{Chung-2018}
J.~S. Chung, A.~Nagrani, and A.~Zisserman,
\newblock ``Voxceleb2: Deep speaker recognition,''
\newblock in {\em Interspeech}, 2018.

\bibitem{He-2015}
Kaiming He, Xiangyu Zhang, Shaoqing Ren, and Jian Sun,
\newblock ``Deep residual learning for image recognition,''
\newblock in {\em ILSVRC}, 2015.

\bibitem{Xie-2019}
W.~Xie, A.~Nagrani, J.~S. Chung, and A.~Zisserman,
\newblock ``Utterance level aggregation for speaker recognition in the wild,''
\newblock in {\em Proc. ICASSP}, 2019.

\bibitem{Zeinali-2019}
Hossein Zeinali, Shuai Wang, Anna Silnova, Pavel Matějka, and Oldřich Plchot,
\newblock ``But system description to voxceleb speaker recognition challenge
  2019,''
\newblock in {\em arXiv:1910.12592}, 2019.

\bibitem{Zhou-2019}
T.~Zhou, Y.~Zhao, J.~Li, Y.~Gong, and J.~Wu,
\newblock ``Cnn with phonetic attention for text-independent speaker
  verification,''
\newblock in {\em IEEE ASRU}, 2019.

\bibitem{Xie-2017}
S.~Xie, R.~Girshick, P.~Dollar, Z.~Tu, and K.~He,
\newblock ``Aggregated residual transformations for deep neural networks,''
\newblock in {\em IEEE CVPR}, 2017.

\bibitem{Hu-2018}
J.~Hu, L.~Shen, and G.~Sun,
\newblock ``Squeeze-and-excitation networks,''
\newblock in {\em IEEE CVPR}, 2018.

\bibitem{Gao-2019}
S.~{Gao}, M.~{Cheng}, K.~{Zhao}, X.~{Zhang}, M.~{Yang}, and P.~H.~S. {Torr},
\newblock ``Res2net: A new multi-scale backbone architecture,''
\newblock {\em IEEE Transactions on Pattern Analysis and Machine Intelligence},
  pp. 1--1, 2019.

\bibitem{Zhou-2020}
T.~Zhou, Y.~Zhao, and J.~Wu,
\newblock ``Resnext and res2net structure for speaker verification,''
\newblock in {\em arXiv:2007.02480}, 2020.

\bibitem{Zhu-2018}
Y.~Zhu, T.~Ko, D.~Snyder, B.~Mak, and D.~Povey,
\newblock ``Self-attentive speaker embeddings for text-independent speaker
  verification,''
\newblock in {\em INTERSPEECH}, 2018.

\bibitem{Okabe-2018}
K.~Okabe, T.~Koshinaka, and Koichi Shinoda,
\newblock ``Attentive statistics pooling for deep speaker embedding,''
\newblock in {\em arXiv:1803.10963}, 2018.

\bibitem{Kye-2020}
S.~M. Kye, Y.~Kwon, and J.~S. Chung,
\newblock ``Cross attentive pooling for speaker verification,''
\newblock in {\em arXiv:2008.05983}, 2020.

\bibitem{Mingote-2019}
V.~Mingote, A.~Miguel, D.~Ribas, A.~Ortega, and E.~Lleida,
\newblock ``Optimization of false acceptance/rejection rates and decision
  threshold for end-to-end text-dependent speaker verification systems,''
\newblock in {\em arXiv:2003.11982}, 2019.

\bibitem{Chung-2020}
J.~S. Chung,
\newblock ``In defence of metric learning for speaker recognition,''
\newblock in {\em arXiv:2003.11982}, 2020.

\bibitem{Shobaki-2000}
Khaldoun Shobaki, John paul Hosom, and Ronald~A. Cole,
\newblock ``The ogi kids’ speech corpus and recognizers,''
\newblock in {\em In ICSLP}, 2000.

\bibitem{Ravanelli-2018}
M.~Ravanelli, T.~Parcollet, and Y.~Bengio,
\newblock ``The pytorch-kaldi speech recognition toolkit,''
\newblock in {\em arXiv:1811.07453}, 2018.

\bibitem{Wang-2018}
H.~Wang, Y.~Wang, Z.~Zhou, X.~Ji, D.~Gong, J.~Zhou, Z.~Li, and W.~Liu,
\newblock ``Cosface: Large margin cosine loss for deep face recognition,''
\newblock in {\em CVPR, 2018, pp. 5265–5274}, 2018.

\bibitem{Yamagishi-2019}
Junichi Yamagishi, Christophe Veaux, and Kirsten MacDonald,
\newblock ``Vcstr vctk corpus: English multi-speaker corpus for cstr voice
  cloning toolkit (version 0.92), [sound],''
\newblock in {\em University of Edinburgh. The Centre for Speech Technology
  Research (CSTR)}, 2019.

\bibitem{Snyder-2015}
D.~Snyder, G.~Chen, and D.~Povey,
\newblock ``Musan: A music, speech, and noise corpus,''
\newblock in {\em arXiv:1510.08484}, 2015.

\bibitem{Ko-2017}
T.~Ko, V.~Peddinti, D.~Povey, M.~L. Seltzer, and S.~Khudanpur,
\newblock ``A study on data augmentation of reverberant speech for robust
  speech recognition,''
\newblock in {\em Proc. ICASSP}, 2017, pp. 5220--5224.

\bibitem{Nagrani-2020}
A.~Nagrani, J.~S. Chung, and A.~Zisserman,
\newblock ``Voxsrc 2020: The second voxceleb speaker recognition challenge,''
\newblock in {\em arXiv:2012.06867}, 2020.

\bibitem{45857}
Jort~F. Gemmeke, Daniel P.~W. Ellis, Dylan Freedman, Aren Jansen, Wade
  Lawrence, R.~Channing Moore, Manoj Plakal, and Marvin Ritter,
\newblock ``Audio set: An ontology and human-labeled dataset for audio
  events,''
\newblock in {\em Proc. IEEE ICASSP 2017}, New Orleans, LA, 2017.

\bibitem{Chen-2020}
H.~Chen, W.~Xie, A.~Vedaldi, and A.~Zisserman,
\newblock ``Vggsound: A large-scale audio-visual dataset,''
\newblock in {\em Proc. IEEE ICASSP 2020}, 2020.

\end{thebibliography}

\end{document}